# FINITE COULOMB CRYSTAL FORMATION

J. Vasut, T.W. Hyde, and L. Barge

*Center for Astrophysics, Space Physics and Engineering Research (CASPER)*
*Baylor University, Waco, TX 76798-7310, USA*

## ABSTRACT

Dust particles immersed within a plasma environment, such as those found in planetary rings or comets, will acquire an electric charge. If the ratio of the inter-particle potential energy to average kinetic energy is large enough the particles will form either a "liquid" structure with short-range ordering or a crystalline structure with long-range ordering. Since their discovery in laboratory environments in 1994, such crystals have been the subject of a variety of experimental, theoretical and numerical investigations. Most numerical and theoretical investigations have examined infinite systems assuming periodic boundary conditions. Since experimentally observed crystals can be comprised of a few hundred particles, this often leads to discrepancies between predicted theoretical results and experimental data. In addition, recent studies have concentrated on the importance of random charge variations between individual dust particles, but very little on the importance of size variations between the grains. Such size variations naturally lead to inter-grain charge variations which can easily become more important than those due to random charge fluctuations (which are typically less than one percent). Although such size variations can be largely eliminated experimentally by introducing mono-dispersive particles, many laboratory systems and all astrophysical environments contain significant size distributions. This study utilizes a program to find the equilibrium positions of a dusty plasma system as well as a modified Barnes-Hut code to model the dynamic behavior of such systems. It is shown that in terms of inter-particle spacing and ordering, finite systems are significantly different than infinite ones, particularly for the most-highly ordered states.

## INTRODUCTION

Complex (dusty) plasma systems are common in a variety of terrestrial and astrophysical environments such as planetary ring systems, cometary environments, planetary nebulae, and plasma processing equipment. Dust particles immersed in a plasma will acquire an electric charge and often have a significant impact on the plasma. The presence of the dust influences the plasma and in some situations can become a component of the plasma, in essence, a giant ion. Since the dust particles have such a large mass and charge as compared to the other plasma species, the resulting "complex" plasma will have several interesting properties.

One of the most interesting these is the possibility of the dust particles arranging themselves into either a well-ordered crystalline state or a "liquid" state with short-range ordering. Such Coulomb crystallization of negatively-charged dust particles was first predicted in 1986 (Ikezi), observed experimentally in 1994 (Thomas, et al., 1994; Chu and I, 1994) and has since been observed in a variety of plasma environments. Most of these have been in laboratory plasma cells on earth, although there have been a limited number of experiments in microgravity conditions as well (Fortov et al., 1998; Morfill et al., 1999).

Until recently, most theoretical and numerical simulations of Coulomb crystallization have considered infinite crystals with periodic boundary conditions. Although this allows the bulk properties of the systems to be investigated, it neglects any edge effects. Since an experimentally observed crystal may typically have only a few hundred to a few thousand particles, such edge effects can be significant. During the past few years small systems have begun to be studied (Lai and I, 1999; Melzer et al, 2001; Joyce et al, 2001; Peeters et al, 2002; Melzer, 2003). Another important feature of finite clusters is that the entire particle cloud may oscillate in response to external conditions (Wang, et al, 2000; Tskhakaya and Shukla, 2001). The majority of theoretical and numerical models have also assumed mono-dispersive particles. Since most experiments use particles with only a small size variation, this is a relatively accurate representation of these systems. However, astrophysical environments, as well as many laboratory ones, have significant size variation within the dust distribution which will likely effect the ordering present in the system.

**METHODS**

Numerical studies of dusty plasma dynamics have been carried out using the *Box_Tree* computer program over the last several years (Vasut and Hyde, 2001; Vasut, 2001; Vasut et al., 2002; Qiao and Hyde, 2002). *Box_Tree*, written by Richardson (1994) and modified by Mathews (2000) uses a Barnes-Hut (1986) algorithm to calculate the inter-particle forces. The Barnes-Hut algorithm calculates the forces from near-by particles directly but uses an approximation to calculate the force from more distant particles, allowing it to scale with particle number as O(N logN). *Box_Tree* allows a large number of parameters to be varied independently and can follow the trajectories of the particles throughout the simulation. Since observed crystals tend to be primarily formed in the horizontal plane due to the presence of gravity and effects such as ion wakes, the simulations were restricted to a horizontal plane. The inter-particle interaction in this plane is given by the shielded Coulomb potential (Konopka et al. 2000)

$$U = \frac{Qe^{-r/\lambda}}{r} \tag{1}$$

where $Q$ is the particle charge, $r$ is the inter-particle spacing and $\lambda$ is the Debye shielding length.

Typically, boundary conditions in *Box_Tree* are provided by the use of periodic boundary conditions, thus allowing bulk effects to be investigated. When modeling finite systems, the periodic boundary conditions are replaced by a highly-charged ring of particles in order to provide confinement.

Size variation within the dust distribution is created by modifying the initial configuration of the particles to include a Gaussian deviate to the size of each particle. The size of the deviate is specified as a user-supplied fraction of the initial particle's radius. Since the particle charge tends to vary linearly with the radius, the charge is also adjusted by the same factor. Particle mass and velocity are also adjusted proportionally to preserve a constant temperature. *Box_Tree* is then run again with, excepting what has just been changed, identical parameters to obtain a new distribution of particles.

In some circumstances, *Box_Tree* calculates more information than is needed. This is most significant when the goal is to simply find the equilibrium position for a series of particles. In such cases, the gas drag option of *Box_Tree* may be used to eliminate the kinetic energy of the particles, allowing them to reach their equilibrium state. Such a process can take a prohibitively long period of time before equilibrium is reached. In such cases another program, named *Equilibrium_Finder,* is used. This program finds the equilibrium position for a set of particles subject to a specified confining potential. It does this by examining the potential at each particle as well as at points immediately adjacent to the particle. The particles are then moved in the direction of the lowest potential. This process is repeated until no particles are moved (or the program reaches a user-supplied maximum number of steps). Since the particles are simply moved without being given a velocity this allows the final equilibrium positions to be quickly reached. The primary danger in this method is the possibility of reaching a local minimum energy state instead of a global one. This danger is reduced by either creating the system in an initially ordered state and/or adding a random noise factor to the particle positions at each step.

**RESULTS**

**One Dimensional Systems**

To examine the simplest case of a finite system, a one-dimensional line was studied using a version of *Equilibrium_Finder*. A number of particles (ranging from 10 to 1000) were uniformly placed 100 units apart along a line with boundary conditions provided by two outside, immobile particles whose charge was arbitrarily set at 100 times the interior particle charge. For this study the Debye length was, again arbitrarily, held at twice the inter-particle separation. (Cases with different values of the Debye length will be the topic of further study.)

The equilibrium state of these systems is based on the interaction of the inward compression from the confinement potential and the mutual repulsion of the particles. The inter-particle separation for the first three runs (particle number of 10, 50 and 100) are shown in Figure 1. A feature common to all of the runs, except for the one with only 10 particles, is the presence of a large central region where the inter-particle separation is nearly constant.

Near the end of the chain the particle spacing increases significantly, especially for the outer-most pair of particles. It is worth noting that, excepting the N=10 case, the number of particles comprising this boundary between the outside of the chain and the interior bulk is around 4 particles. In every case the outermost four particle pairs have a separation that is at least 1% greater than the minimum particle separation while the 5$^{th}$ pair of particles has a separation within 1% of the minimum. This seems to suggest that (at least for systems up to 1000 particles) the "skin" of such systems is only 4 particles deep. The amount of "stretching" of the end of the chain, defined here as the ratio of inter-particle separation for the end particles as compared to the minimum separation, is nearly constant for N≥50 but does increase slowly with increasing numbers of particles, as shown in Table 1.

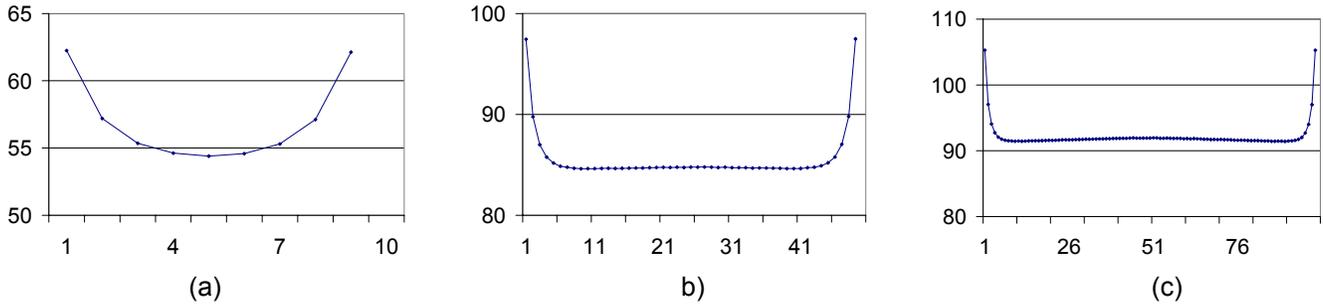

Fig 1. Inter-particle separation for one-dimensional systems with (a) 10, (b) 50 and (c) 100 particles.

Table 1. Increase in inter-particle separation for end particles compared to the minimum separation.

| Number of Particles | 10 | 50 | 100 | 250 | 500 | 1000 |
|---|---|---|---|---|---|---|
| Variation | 14.4% | 15.2% | 15.2% | 15.3% | 15.3% | 15.4% |

**Two Dimensional Systems**

Two dimensional systems were examined using a combination of *Equilibrium_Finder* and *Box_Tree*. *Equilibrium_Finder* was used when only the final, equilibrium position was desired and *Box_Tree* was used when dynamic systems (such as oscillating systems) or systems at finite temperatures were examined.

Edge Effects

To study the edge effects of two dimensional systems, *Equilibrium_Finder* first generated an ordered lattice and then allowed it to either relax or expand naturally to its equilibrium position. An example of the final state for a system composed of 2611 particles is shown in Figure 2(a), with an enlarged view of one quadrant shown in Figure 2(b).

The most important effect which can be seen in Figure 2 is the transition from hexagonal symmetry in the interior of the lattice to cylindrical symmetry near the boundary as imposed by the system boundary conditions. As can be seen, to a large degree, the outer-two most rings have cylindrical symmetry while the interior is solidly hexagonal with a transition layer several layers thick between the two, consistent with results from other authors (Joyce, et al, 2001; Peeters et al, 2002)

Examination of different simulations reveals that the "thickness" of the intermediate layers between the hexagonal interior and the circular exterior is not constant and can range from 3 to 8 layers. This thickness increases with the number of particles in the system; the exact relationship between the two will be the subject of additional study.

The transition between the two types of symmetry naturally produces a breakdown in ordering within the system as can be seen in the pair correlation function for the interior and exterior particles. The dividing line between the two was chosen here to be the distribution radius divided by 0.7071, thus half of the particles will be in each section. As expected, the interior, with its greater symmetry, has an strongly-peaked correlation function, showing a true hexagonal lattice. The outer half of the system is also strongly-peaked, but to a significantly less degree than. This can be seen in Figure 3 which shows the marked difference between the two halves.

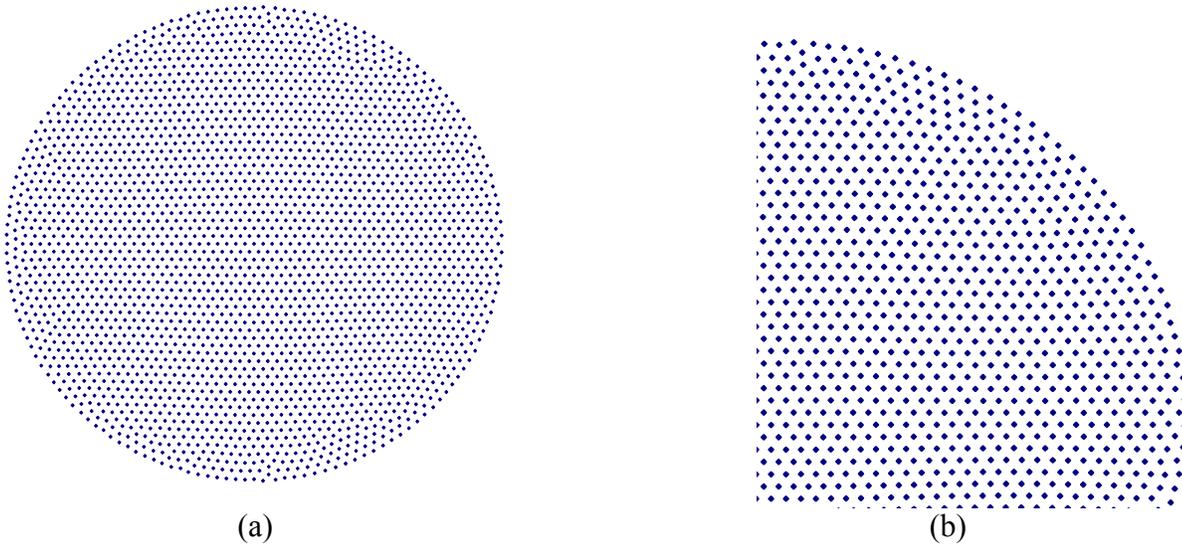
(a)                                   (b)

Fig. 2. Two-dimensional lattice with 2611 particles under cylindrical boundary conditions. Figure (a) is the entire crystal while (b) is the upper-right quadrant at a higher magnification.

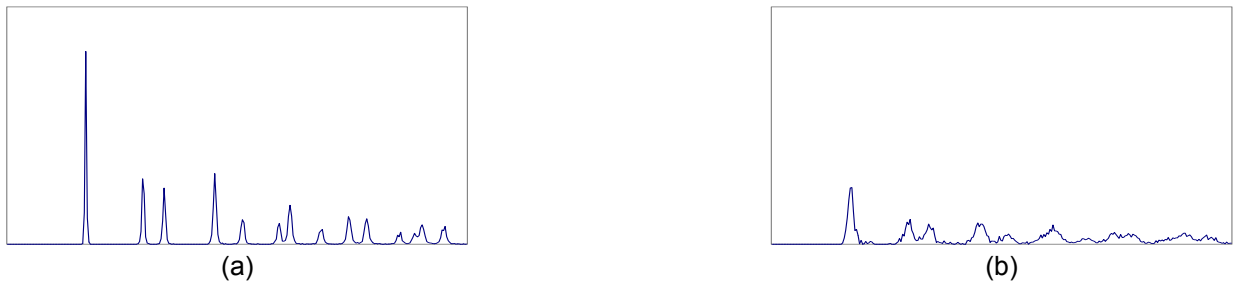
(a)                                   (b)

Fig. 3. Pair correlation function for the (a) interior and (b) exterior of a finite lattice.

Since the number of particles in each shell is dependent upon the square of the radius of that shell, a significant percentage of the particles will be part of the non-hexagonal boundary and/or the transition between the boundary and the interior. In the system shown in Figure 2, comprised of roughly 26 shells, the outer-most two are clearly circular and the transition layer is at least 4 to 5 shells thick. This means that roughly 15% of the particles are in the circular region and 26-32% are in the transition region, leaving between only 53 to 59% in the interior.

Dusty Plasma Oscillations

In addition to finding the equilibrium positions of a finite system of dust particles at zero temperature, oscillations of dust particle systems are also of interest. In such cases the dynamics of the system are important and *Box_Tree* must be used. In these simulations, the particles are initially placed within a circle with some given radius. The particle cloud will then either collapse down to a smaller radius or expand to a larger one. Due to the particles' inertia they generally over-shoot their equilibrium state and then proceed to oscillate about that position. Typically such large-scale oscillations are not observed since Coulomb crystals normally come to an equilibrium condition gradually as the particles settle into the lattice with their excess kinetic energy absorbed by the effects of neutral gas drag. Such oscillations could occur, however if the bias voltage on the electrode was changed suddenly. A similar change could occur in planetary ring systems as portions of the ring enter and leave the planetary shadow, thus changing the in-situ plasma conditions.

The rate of oscillations of the particle cloud depend upon a variety of parameters, ranging from the particle mass and charge to the strength of the confining field and/or the Debye length (Tskhakaya and Shukla, 2001). In this study only the effect of particle charge and mass were studied. The data for this series of runs is shown in Figure 5. A best fit to this data yields a dependence of

$$f = q^{0.52} m^{-0.50} \qquad (2)$$

which is in good agreement with Tskhakaya and Shukla (2001). The slight difference in the charge dependence is primarily due to the fact that the confining potential is not harmonic.

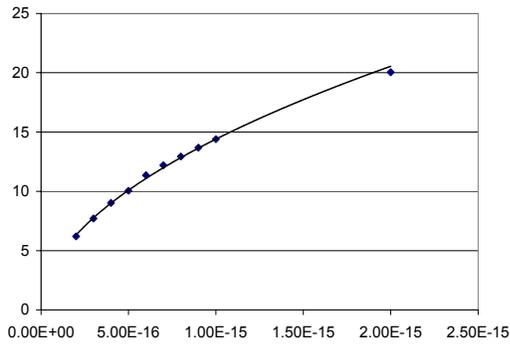 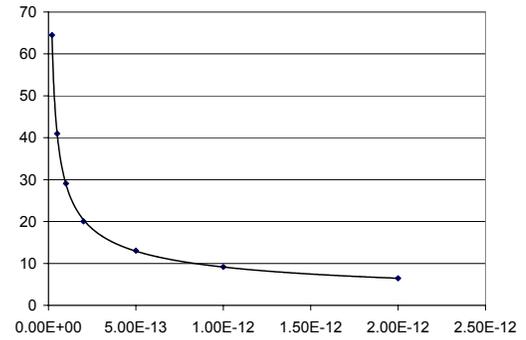

Fig. 5.  Frequency dependence upon (a) particle charge and (b) particle mass.

Effect of Size Variation on Coulomb Crystallization

In most astrophysical environments there will be a significant size variation among the particles. This is also true in many laboratory systems as well. Typically in laboratory environments this variation is minimized, but many systems still have significant size distributions (Smith, et al, 2002). For large size variations the larger (and more highly charged) particles will tend to move toward the center of the particle distribution. If, however, the charge variation is small and/or the system becomes ordered quickly, it is possible that such particles will reach a local equilibrium where the size variation is present randomly through the system. In this study it is assumed that the size/charge distribution is indeed random and follows a Gaussian distribution.

As seen in Figure 6, the charge variation has the effect of damaging the hexagonal structure of the system, an effect which can be most clearly seen in the correlation functions. The correlation function for the case with mono-dispersive particles as well as a 15% size variation is shown in Figure 6(a) where the particles have no kinetic energy and are therefore very highly-ordered. In such cases the size variation makes a large impact on the ordering present, but leaves the system in a very highly ordered state. At finite temperatures, such as the one shown in Figure 6(b), the system is not nearly as ordered and the size variation, while making a significant impact, is less important.

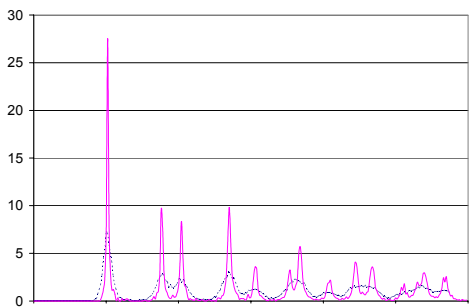 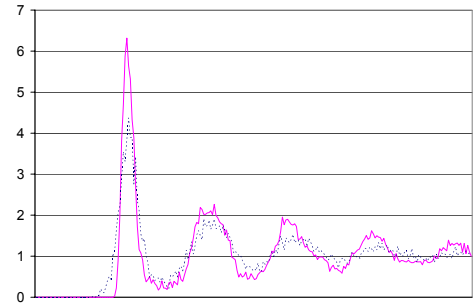

Figure 6.  Pair correlation function with size/charge variation of 0% and 15% (dashed line) for (a) zero temperature and (b) finite temperature.

**CONCLUSIONS**

Finite dusty plasma systems have been shown to have significant edge effects, consistent with results from other authors (Joyce et al, 2001; Peeters, et al 2002). In one dimension these are largely limited to the four outer-most pairs of particles having significantly larger inter-particle separation than those in the bulk of the system. This effect depends only very weakly upon the number of particles (at least when the number of particles is more than 10 or so). In two dimensions the edge of the crystal is forced into a circular shape by the confining potential. Since the interior of the system has hexagonal symmetry this creates a layer several particles thick where the exterior and interior come together in a relatively disordered state (when compared to the order present in the interior).

Finite systems may also oscillate if the system is not at an equilibrium state. Such oscillations could easily be induced by a rapid change of the plasma and/or confining potential. In such cases the dust distribution will oscillate with a frequency that closely depends on $\sqrt{q/m}$. Such results are again consistent with other work (Tskhakaya and Shukla, 2001)

Dusty plasma systems may also have a size variation, especially astrophysical ones. Such size variation lead naturally to a charge variation which can have a significant impact on the amount of ordering present within the system. The size of this impact is greatest at zero temperatures where there is naturally more order present than at finite temperatures.

Email address of T.W. Hyde: Truell_Hyde@baylor.edu